\documentclass[12pt,preprint]{aastex}
\usepackage{mathrsfs}
\usepackage{color}

\usepackage{amsmath}

\usepackage{graphics}

\slugcomment{revised submitted to ApJ, on \today}

\shorttitle{What governs the bulk velocity of the jet components in
active galactic nuclei?}

\shortauthors{Chai, Cao \& Gu}

\begin{document}

\title{What governs the bulk velocity of the jet components in active galactic
nuclei? }

\author{Bo Chai, Xinwu Cao and Minfeng Gu}
\affil{Key Laboratory for Research in Galaxies and Cosmology,
Shanghai Astronomical Observatory, Chinese Academy of Sciences, 80
Nandan Road, Shanghai, 200030, China;\\
chaibo@shao.ac.cn;cxw@shao.ac.cn;gumf@shao.ac.cn}

\begin{abstract}
We use a sample of radio-loud active galactic nuclei (AGNs) with
measured black hole masses to explore the jet formation mechanisms
in these sources. Based on the K\"{o}nigl's inhomogeneous jet model,
the jet parameters, such as the bulk motion Lorentz factor, magnetic
field strength, and electron density in the jet, can be estimated
with the very long-baseline interferometry and X-ray data. We find a
significant correlation between black hole mass and the bulk Lorentz
factor of the jet components for this sample, while no significant
correlation is present between the bulk Lorentz factor and the
Eddington ratio. The massive black holes will be spun up through
accretion, as the black holes acquire mass and angular momentum
simultaneously through accretion. Recent investigation indeed
suggested that most supermassive black holes in elliptical galaxies
have on average higher spins than the black holes in spiral
galaxies, where random, small accretion episodes (e.g., tidally
disrupted stars, accretion of molecular clouds) might have played a
more important role. If this is true, the correlation between black
hole mass and the bulk Lorentz factor of the jet components found in
this work implies that the motion velocity of the jet components is
probably governed by the black hole spin. No correlation is found
between the magnetic field strength at $10R_{\rm S}$ ($R_{\rm
S}=2GM/c^2$ is the Schwarzschild radius) in the jets and the bulk
Lorentz factor of the jet components for this sample. This is
consistent with the black hole spin scenario, i.e., the faster
moving jets are magnetically accelerated by the magnetic fields
threading the horizon of more rapidly rotating black holes. The
results imply that the Blandford-Znajek (BZ) mechanism may dominate
over the Blandford-Payne (BP) mechanism for the jet acceleration at
least in these radio-loud AGNs.
\end{abstract}

\section{Introduction}

Only a small fraction of active galactic nuclei (AGNs) are
radio-loud \citep*[e.g.,][]{1989AJ.....98.1195K}. Relativistic jets
have been observed in many radio-loud AGNs, which are believed to be
formed very close to black holes. The power of jets is supposed to
be extracted from the accretion disk or the rotating black hole. The
currently most favored models of jet formation are Blandford-Znajek
(BZ) and Blandford-Payne (BP) mechanisms
\citep{1977MNRAS.179..433B,1982MNRAS.199..883B}. In the BZ process,
energy and angular momentum are extracted from a rotating black hole
and transferred to a remote astrophysical load by open magnetic
field lines. In the BP process, the magnetic field threading the
disk extracts energy from the rotating gas in the accretion disk to
power the jet/outflow. The relative importance of these two
mechanisms has been extensively explored by many different authors,
which is still debating
\citep*[e.g.,][]{1996MNRAS.283..854M,1999ApJ...512..100L,2011ApJ...737...94C}.

The apparent motions of the jet components in AGNs were detected by
multi-epoch very long-based interferometry (VLBI) observations
\citep*[e.g.,][]{2004ApJ...609..539K,2005AJ....130.1389L}. The
Lorentz factor and the viewing angle of the jet component can be
derived with the measured proper motion of the jet component if the
Doppler factor is estimated. There are several different approaches
applied to estimate the Doppler factor of the jets.
\citet{1994ApJ...426...51R} estimated the equipartition Doppler
boosting factor assuming that the sources are in equipartition
between the energy of radiating particles and the magnetic field in
the jets \citep*[also see][]{1996ApJ...461..600G}. The variability
Doppler factor is derived on the assumption that the associated
variability brightness temperature of total radio flux density
flares are caused by the relativistic jets
\citep{1999ApJ...521..493L}. Based on the synchrotron self-Compton
(SSC) model, the physical quantities in the jets can be estimated by
using the VLBI observations and the X-ray flux density on the
assumption of homogeneous spherical emission plasma
\citep{1987slrs.work..280M,1993ApJ...407...65G}. The inhomogeneous
relativistic jet model can reproduce both the flat spectrum
characteristics of some AGNs and the dependence of the core size on
the observing frequency
\citep{1979ApJ...232...34B,1981ApJ...243..700K}. Based on this
inhomogeneous jet model, an approach was suggested by
\citet{1998ApJ...494..139J} to estimate the jet parameters including
bulk Lorentz factor $\Gamma$, viewing angle $\theta$, and electron
number density $n_{\rm e}$ in the jets of AGNs.

The relation between the jets and the accretion disks were
extensively explored in many previous works
\citep*[e.g.,][]{1989MNRAS.240..701R,1997MNRAS.286..415C,1999MNRAS.307..802C,2001MNRAS.320..347C,2009MNRAS.396..984G},
which indicate the jets are indeed closely linked to the accretion
disks, though the different jet formation mechanisms are still
indistinguishable. The relationship between black hole mass and the
motion of the jet components for a sample of blazars with measured
proper motion of jet components by multi-epoch VLBI observations was
investigated by \citet{2009RAA.....9..293Z}. They found a
significant intrinsic correlation between black hole masses and the
minimal Lorentz factors of the jet components, while the Eddington
ratio is only weakly correlated with the minimal Lorentz factor.
They suggested that the BZ mechanism may dominate over the BP
mechanism for the jet acceleration at least in these blazars, if
massive black holes are spinning more rapidly than their less
massive counterparts.

In this work, we use a sample of radio-loud AGNs with jet parameters
estimated with the inhomogeneous jet model to re-investigate the
relationship of the Lorentz factor of jets with black hole mass, the
Eddington ratio, or the strength of the magnetic field in the jets.
The sample and the physical parameters used in this paper are
described in Section 2. Section 3 contains the results. The last
section includes the discussion. The cosmology with $H_{0}=70 \rm
{~km ~s^ {-1}~Mpc^{-1}}$, $\rm \Omega_{M}=0.3$, and $\rm
\Omega_{\Lambda} = 0.7$ have been adopted throughout this work.

\section{Estimate of the jet parameters}

Based on the K\"{o}nigl's inhomogeneous jet model, the jet
parameters including the bulk Lorentz factor $\Gamma$, viewing angle
$\theta$, and electron number density $n_{\rm e}$ in the jets can be
estimated with the data of VLBI and X-ray observations. We summarize
the inhomogeneous jet model and the approach used to estimate the
jet parameters in this section \citep*[see][for the
details]{1998ApJ...494..139J,2009MNRAS.396..984G}.


K\"onigl (1981) constructed an inhomogeneous jet model, in which the
magnetic field strength and number density of the relativistic
electrons are assumed to vary with the distance from the
apex of the jet $r$ as $B(r)={B_1}(r/{r_1})^{-m}$ and ${n_{\rm e}}(r,{\gamma _{\rm e}})={n_1}%
(r/{r_1})^{-n}{\gamma _{\rm e}}^{-(2\alpha +1)}$ in the jet,
respectively, where ${r_1}=1$ pc and $\gamma_{\rm e}$ is the Lorentz
factor of the electron in the jet. In this jet model, the conical
jet with a half opening angle $\phi$ is moving in the direction at a
viewing angle of $\theta$ with respect to line of sight. The
distance $r({\tau _{\nu _{\rm s}}}=1)$, at which the optical depth
to the synchrotron self-absorption at the observing frequency $\nu
_{\rm s}$ equals unity, is given by
\begin{equation}
{\frac{{r({\tau_{\nu_{\rm s}}}=
1)}}{{r_1}}}=(2c_{2}(\alpha)r_{1}n_{1}\phi\csc
\theta)^{2/(2\alpha+5)k_{\rm
m}}(B_{1}\delta)^{(2\alpha+3)/(2\alpha+5)k_{\rm m}}(\nu_{\rm s}
(1+z))^{-1/k_{\rm m}}, \label{r_optthick}
\end{equation}
where $c_{2}(\alpha)$ is the constant in the synchrotron absorption
coefficient, $\delta$ is the Doppler factor, and $k_{\rm
m}=[2n+m(2\alpha+3)-2] /(2\alpha+5)$.

The optically thick VLBI core size corresponds to the projection of
the distance $r({\tau _{\nu _{\rm s}}}=1)$, and therefore the VLBI
core angular size $\theta_{\rm d}$ is
\begin{equation}
\theta_{\rm d}= {\frac{{r({\tau_{\nu_{\rm s}}}= 1)\sin \theta}}{{\
D_{\rm a}}}}, \label{theta_d}
\end{equation}
where $D_{\rm a}$ is the angular diameter distance of the source.

By integrating the observed intensity over the projected area of the
jet, the total optically thick flux is
\begin{equation}
s(\nu_{\rm s})={\frac{{r_{1}^{2}\phi\sin\theta}}{{(4+m)\pi D_{\rm
a}^{2}}}} {\frac{{\ c_{1}(\alpha)}}{{c_{2}(\alpha)}}}
B_{1}^{-1/2}\nu_{\rm s}^{5/2} \left({\frac{
\delta}{{1+z}}}\right)^{1/2}\left({\frac{{r({\tau_{\nu_{\rm s}}}=
1)}}{{r_1}}} \right)^{(4+m)/2}, \label{s_core}
\end{equation}
where $\nu _{\rm s}$ is the VLBI observing frequency, and
${c_1}(\alpha )$ and ${c_2}(\alpha )$ are the constants in the
synchrotron emission and absorption coefficients, respectively. The
relation between apparent transverse velocity $\beta _{\rm app}$ and
the bulk velocity of the jet ${\beta }c$ is
\begin{equation}
\beta_{\rm app}={\frac{{\beta\sin\theta}}{{1-\beta\cos\theta}}}.
\label{beta_app}
\end{equation}

The X-ray flux density of the unresolved jet can be calculated with
Equation (13) in \citet{1981ApJ...243..700K}'s work. The frequency
region ${\nu _{\rm c}}>{\nu _{\rm cb}}(r_{\rm M})$ was adopted,
where $r_{\rm M}$ is the smallest radius from which optically thin
synchrotron emission with spectral index $\alpha $ is observed
\citep*[see][for the details]{1981ApJ...243..700K}.

Given the three parameters, $\alpha $, $m$, and $n$, the four
independent variables, $n_1$, $B_1$, $\beta$, and $\theta$ can be
derived from Equations (\ref{theta_d})-(\ref{beta_app}) together
with \citet{1981ApJ...243..700K}'s equation (13) by using the data
of the VLBI and X-ray observations.  The total electron number
density $n_{\rm t}$ is given by
\begin{equation}
n_{\rm t}=\int\limits_{\gamma_{\rm min}}^{\gamma_{\rm max}}n_{\rm
e}(r, \gamma_{\rm e}) d\gamma_{\rm e}.
\end{equation}
The model parameters, $\alpha=0.75$, $m=1$ and $n=2$ are adopted in
\citet{2009MNRAS.396..984G} for a jet with mass conserved along $r$.
{The kinetic power of jet can be calculated with
\begin{equation}
L_{\rm kin} = \frac{4}{3}\pi r_{1}^{2}n_{1}\gamma_{\rm
e,min}^{-\frac{3}{2}} [1-\cos({1/\Gamma})]m_{\rm
e}\langle\gamma_{\rm e}\rangle\Gamma(\Gamma-1)\beta c^{3},
\label{l_kin}
\end{equation}
for electron-positron jets, and
\begin{equation}
L_{\rm kin} = \frac{4}{3}\pi r_{1}^{2}n_{1}\gamma_{\rm
e,min}^{-\frac{3}{2}} [1-\cos({1/\Gamma})](m_{\rm
e}\langle\gamma_{\rm e}\rangle+m_{\rm p}\langle\gamma_{\rm
p}\rangle)\Gamma(\Gamma-1)\beta c^{3}, \label{l_kin_ep}
\end{equation}
for electron-proton jets, where $\Gamma$ is the Lorentz factor of
the jet, $1/\Gamma$ is the half opening angle of the conical
jet, $m_{\rm e}$ is the electron rest mass, $m_{\rm p}$ is the rest
mass of proton, $\langle\gamma_{\rm e}\rangle$ is the average
Lorentz factor of electrons, and $\langle\gamma_{\rm p}\rangle$ is
the average Lorentz factor of protons. We have assumed the
positrons have the same energy distribution as the electrons in
electron-positron jets, which contribute a half portion of the
observed emission from the jets. This means that the density $n_e$
derived from the observational data based on the inhomogeneous jet
model is the total number density of the electrons and positrons in
the jets.

The composition of the jet plasma is still an unresolved issue. The
\citet{1996MNRAS.283..873R} analyzed VLBI data of M87 and concluded
that the core is probably dominated by electron-positron plasma. By
detecting circular polarization, \citet{1998Natur.395..457W}
suggested that extragalactic radio jets are composed mainly of
electron-positron plasma with $\gamma_{\textrm{e,min}}\la 10$.
Considering the dynamic and radiation properties,
\citet{2004MNRAS.349..336K} and \citet{2005ApJ...619...73H}
suggested that the sources they studied are composed of
electron-positron plasma. The presence of protons and the
minimal energy of electrons in the hot spots of the radio lobes are
constrained by multi-waveband observations
\citep*{2006ApJ...644L..13B,2007ApJ...662..213S,2009ApJ...695..707G}.
\citet{2000ApJ...534..109S} suggested that the X-ray spectral observations in OVV quasars require the
composition of the jets to be a mixture of electrons-positrons and
protons. The detailed calculations of the pressure of the cocoon in
Cygnus A did not give a tight constraint on the jet composition,
i.e., either electron-positron or electron-proton plasma is possible
in the jets of this source \citep*{2012ApJ...751..101K}. Faraday
rotation is sensitive to the plasma composition, which can also be
used to constrain the composition of jets
\citep*[e.g.,][]{2010MNRAS.403.1993P}. It was found that the
presence of protons is required to explain the observed circular
polarization and Faraday rotation in radio cores of blazars
\citep*[e.g.,][]{2008MNRAS.391..124V,2010MNRAS.403.1993P}. For
electron-proton jets, $\gamma_{\rm e,min}\sim100$ is suggested, and
the low cut-off energy of jet can be as low as unity for
electron-positron jets \citep{1993MNRAS.264..228C}, while
\citet{1998Natur.395..457W} suggested that the jets are
electron-positron plasmas with $\gamma_{\rm e,min}\la 10$ at least
in some sources. For electron-proton jets, assuming the inverse
Compton scattering origin of X-ray emission from PKS 0637$-$752,
\citet{2000ApJ...544L..23T} find that, $\gamma_{\rm e,min}=10$, if
the seed photon come from radiation field external to the jets
(e.g., the cosmic microwave background), or $\gamma_{\rm
e,min}=2\times10^3$ for the sychrontron self-Compton case. We note
that only the normalization is changed if the different plasma
composition is considered, which means that most of the statistic
results derived in this work is independent of the jet composition.
The kinetic power derived for electron-positron pair plasma is
roughly consistent with that derived for electron-proton plasma with
conventionally used minimal values of electron energy. The bulk
kinetic power $L_{\rm kin}$ for an electron-proton jet with
$\gamma_{\rm e,min}=100$ is in agreement with that of
electron-positron jets with $\gamma_{\rm e,min}=10$ within a factor
of 3, as $m_{\rm p}=1836m_{\rm e}$ and $\langle\gamma_{\rm
p}\rangle=1$ for electron-proton jets \citep{2009MNRAS.396..984G}.
We find that the kinetic luminosity $L_{\rm kin}$ will be reduced by
about a factor of 3 if $\gamma_{\rm e,min}$ is changed from $10$ to
$1$. The choice of matter composition of jets does not change the
bulk kinetic power of jet significantly, and most of the statistic
analyses are not affected by the matter composition of jets. For
simplicity, we therefore calculate the bulk kinetic power $L_{\rm
kin}$ assuming electron-positron jets with
$\gamma_{\textrm{e,min}}=10$ in this work.

\section{The sample}

\citet{2009MNRAS.396..984G} constructed a sample of 128 sources, of
which the jet parameters are derived from the VLBI and X-ray data
with the K\"{o}nigl's inhomogeneous jet model. We search the
literature for the black hole mass measurements, and finally obtain
a sample of 101 sources with measured black hole masses, including
77 quasars, 20 BL Lac objects and 4 radio galaxies. The black hole
masses for most sources in our sample are estimated by using an
empirical relation between BLR size and ionizing luminosity together
with measured broad-line widths assuming the BLR clouds being
gravitationally bound by the central black hole
\citep*[e.g.,][]{2003ApJ...583..124S,2006ApJ...637..669L,2008ApJ...680..169S,2009MNRAS.398.1905W,2009RAA.....9..293Z}.
For some sources, especially BL Lac objects or radio galaxies, the
black hole masses can be estimated from the properties of their host
galaxies with either $M_{\rm BH}$-$\sigma$ or $M_{\rm BH}$-$L$
relations, where $\sigma$ and $L$ are the stellar velocity
dispersion and the bulge luminosity of the host galaxies
\citep*[e.g.,][]{2005ApJ...631..762W,2003ApJ...599..147C}. The black
hole masses can also be estimated with the $\gamma$-ray variability
timescales for some $\gamma$-ray blazars
\citep{2003MNRAS.340..632L}. For a few sources, the black hole
masses are estimated from reverberation mapping
\citep{2004ApJ...613..682P} and stellar kinetics
\citep{2006A&A...455..173P}(see Table 1). The bolometric luminosity
$L_{\rm bol}$ is estimated from the total broad-line luminosity by
assuming $L_{\rm bol}=10L_{\rm BLR}$ \citep{1990agn..conf...57N}.
All the data for the sample are summarized in Table 1, in which all
the jet parameters are taken from \citet{2009MNRAS.396..984G}.



\section{Results}

We plot the relation between black hole mass and the bulk Lorentz
factor of the jet components in Figure \ref{mbh_gamma}. {A strong
correlation is found between these two quantities with a Spearman
rank correlation coefficient $r=0.357$ at 99.98 percent confidence
level. Using the linear regression analysis, the correlation can be
expressed as
\begin{equation}
\log \Gamma = (0.20\pm0.06) \log M_{\rm BH}-(0.68\pm0.52),
\end{equation}
and it becomes
\begin{equation}
\log \Gamma = (0.21\pm0.07) \log M_{\rm BH}-(0.70\pm0.61),
\end{equation}
for the quasars in the sample with a correlation coefficient
$r=0.376$ at 99.92 percent confidence level.}

In Figure \ref{mbh_gamma_z}, we plot the relation between black hole
mass and redshift $z$, and the relation between the bulk Lorentz
factor of the jet components and redshift $z$. It is found that both
the black hole mass and the Lorentz factor are strongly correlated
with redshift $z$, which implies that the correlation between black
hole mass and the bulk Lorentz factor of the jet components may be
caused by the common dependence of redshift. We therefore choose a
sub-sample of the sources in a restricted range of redshift
$0.8<z<1.2$ (see the sources between two vertical dotted lines in
Figure \ref{mbh_gamma_z}). No significant correlations are present
between $M_{\rm BH}$ and $z$, or $\Gamma$ and $z$, while a
significant correlation between $M_{\rm BH}$ and $\Gamma$ is still
present for the sources in this sub-sample (see Figure
\ref{mbh_gamma_sub} and Table 2). Therefore, we conclude that the
significant correlation between black hole mass and the bulk Lorentz
factor might be intrinsic, at least for our present sample, which is
confirmed by the partial correlation analyses (see Table 2).

The relation between the Eddington ratio and the bulk Lorentz factor
of the jet components is plotted in Figure \ref{eddratio_gamma}.
{The correlation analysis shows that no significant correlation is
found between $L_{\rm bol}/L_{\rm Edd}$ and $\Gamma$ with a
correlation coefficient $r=0.099$ at 63.94 percent confidence
level.}

As the masses of the black holes in this sample are in the range of
$\sim 10^{7}-10^{10}M_\odot$, we plot the relation of the bulk
Lorentz factor of jet $\Gamma$ with the magnetic field strength at
$10R_{\rm S}$ in Figure 5. {No correlation is found between these
two quantities with a correlation coefficient $r=0.01$ at 7.78
percent confidence level. We summarize the statistic results in
Table 2.}

We define jet efficiency $\eta_{\rm jet}$ as
\begin{equation}
L_{\textrm{kin}}=\eta_{\rm jet}\dot{M}_{\rm acc}c^2,  \label{L eta}
\end{equation}
where $\dot{M}_{\rm acc}$ is the mass accretion rate. The mass
accretion rate can be estimated from the bolometric luminosity,
i.e., $L_{\rm bol}=0.1\dot{M}_{\rm acc}c^2$, and therefore,
\begin{equation}
\eta_{\rm jet}={\frac {0.1L_{\rm kin}}{L_{\rm bol}}}.  \label{jet efficiency}
\end{equation}
The distribution of the jet efficiency for our sample is given in
Figure \ref{eta number}. It is found that the efficiencies
of some sources are significantly greater than unity, which may be
due to a fixed $\gamma_{\rm e,min}$ for all sources. The possibility
of different values of $\gamma_{\rm e,min}$ in some individual
sources cannot be ruled out (see the discussion in Section 2). The
relation between the bolometric luminosity and the kinetic power of
the jet is plotted in Figure \ref{jet efficiency}. The
magnetic energy density in the jets can be expressed as $U_{\rm
B}=B^2/8\pi$, and the magnetic energy flux in the jets can be
calculated with $B_{1}^2r_{1}^2c[1-\rm{cos}(1/\Gamma)]\Gamma^2/4$. The ratio
of the bulk kinetic power $L_{\rm kin}$ to the magnetic energy flux
$L_{\rm B}$ in the jets is plotted in Figure \ref{l_kin l_b}.


\section{Discussion}

We find an intrinsic correlation between the black hole masses and
the Lorentz factors of the jet components for a sample of radio-loud
AGNs, while no significant correlation between the Eddington ratios
and the Lorentz factors is present for the same sample.  No
correlation is found between the magnetic field strength at a given
distance in units of Schwarzschild radius in the jet and the bulk
Lorentz factor. These results provide useful clues on the mechanisms
of jet formation and acceleration in radio-loud AGNs.

Mass accretion in the AGN phases plays a dominant role in the growth
of massive black holes in the centers of galaxies
\citep*[e.g.,][]{1982MNRAS.200..115S,2002ApJ...574..740T}. As
massive black holes acquire mass and angular momentum simultaneously
through accretion, the black holes will be spun up with mass growth.
The mergers of black holes may also affect the spin evolution of
massive black holes \citep{2003ApJ...585L.101H}.
\citet{2007ApJ...667..704V} investigated how the accretion from a
warped disc influences the evolution of black hole spins with the
effects of accretion and merger being properly considered and
concluded that within the cosmological framework, most supermassive
black holes in elliptical galaxies have on average higher spins than
black holes in spiral galaxies, where random, small accretion
episodes (e.g., tidally disrupted stars, accretion of molecular
clouds) might have played a more important role.
\citet{2008MNRAS.390..561C} calculated the black hole mass function
of AGN relics with the observed Eddington ratio distribution of
AGNs, and compared it with the measured mass function of the massive
black holes in galaxies. They found that the radiative efficiencies
of most massive accreting black holes are higher than those of less
spinning black holes. If this is the case, the correlation between
$M_{\rm BH}$ and $\Gamma$ found in this work indicates that the bulk
velocity of jets is mainly regulated by the black hole spin
parameter $a$. The kinetic power of the jet is,
\begin{equation}
L_{\rm kin}=\Gamma \dot{M}_{\rm jet}c^2, \label{l_kin_1a}
\end{equation}
where $\dot{M}_{\rm jet}$ is the mass loss rate in the jet. If the
jet is powered through the BZ process, we have
\begin{equation}
L_{\rm kin}\sim L_{\rm BZ}\propto M_{\rm BH}a^2,
\end{equation}
for a radiation pressure dominated accretion disk surrounding a
rotating black hole \citep{1997MNRAS.292..887G}, where $a$ is the
black hole spin parameter. Combining Equations (11) and (12), we
have
\begin{equation}
\Gamma\propto \left({\frac {\dot{M}_{\rm jet}}{M_{\rm
BH}}}\right)^{-1}a^2.
\end{equation}
This indicates that the Lorentz factor of the jet $\Gamma$
should be anti-correlated with $\dot{M}_{\rm acc}/M_{\rm BH}$ if
$\dot{M}_{\rm jet}\propto \dot{M}_{\rm acc}$, which seems to be
inconsistent with our statistic results. It implies that such a jet
formation model for a radiation pressure dominated accretion disk
may not be applicable for the blazars considered in this work. In
fact, the structure of the radiation pressure dominated accretion
disks should be altered significantly in the presence of strong
magnetic fields \citep{2012ApJ...753...24L}, which was not
considered in \citet{1997MNRAS.292..887G}.

Recent numerical simulations show that an accretion flow can
evolve into a magnetically arrested flow, and at this state the
outflow efficiency can be extremely high (as high as $\ga 100$\%)
\citep{2011MNRAS.418L..79T,2012arXiv1201.4163M,2012MNRAS.tmpL.445T}.
The strength of the field in the magnetically arrested accretion
flow can be estimated as
\citep{2011MNRAS.418L..79T,2012arXiv1201.4163M},
\begin{equation}
B^{2}\propto \frac{\dot{M}_{\textrm{acc}}}{M_{\rm BH}^{2}}. \label{B}
\end{equation}
If the jet is powered through the BZ process, we have
\begin{equation}
L_{\rm kin}\sim L_{\rm BZ}\propto M_{\rm BH}^{2}B^{2}a^{2},
\label{L BZ}
\end{equation}
where $a$ is the black hole spin parameter. Combining Equations
(\ref{B}) and (\ref{L BZ}), we obtain
\begin{equation}
L_{\rm kin}\propto \dot M_{\rm acc} a^2, \label{l_kin_1b}
\end{equation}
which implies
\begin{equation}
\eta_{\rm jet}\propto a^2.
\end{equation}
Hence, the efficiency of the jet production does not depend on the
accretion rate of the disk. Substitute Equation (\ref{l_kin_1b})
into (\ref{l_kin_1a}), we have
\begin{equation}
\Gamma\propto \frac{\dot M_{\rm acc}}{\dot M_{\rm jet}}a^2\propto
a^2,\label{gamma_a}
\end{equation}
if the mass loss rate in the jets $\dot{M}_{\rm jet}$ is assumed to
be proportional to the mass accretion rate $\dot{M}_{\rm acc}$. The
correlation between black hole mass and the bulk Lorentz factor of
the jet components found in this work implies that the motion
velocity of the jet components is probably governed by the black
hole spin, if the massive black holes are spinning more rapidly than
the less massive counterparts. This is consistent with the
magnetically arrested accretion flow model (see Equation
\ref{gamma_a}). The bulk Lorentz factor of the jets is predicted to
be independent of the accretion rate in this model, which is also
consistent with our statistic results (see Figure
\ref{eddratio_gamma}).

In Figure \ref{l_kin l_b} we find that the magnetic energy
flux is far less than bulk kinetic power. As discussed in Section 2,
the kinetic luminosity $L_{\rm kin}$ will change with a factor of 3
if an electron-proton jet with $\gamma_{\rm min}=100$ is assumed, so
the kinetic energy flux always dominates over magnetic energy flux
in the jets of this blazar sample.

In inhomogeneous jet model, the relativistically
moving plasma expands uniformly, which suffers adiabatic losses. The
electron energy in the jet at parsec scales requires electron
acceleration, otherwise huge energy of electrons is required at the
base of jet. The magnetic field in the jet may be the energy
reservoir for electrons. We see in figure \ref{l_kin l_b} that a few
sources have magnetic field energy larger than the kinetic energy of
particles. However, in either case of the matter content, most of
the sources have kinetic energy of electrons exceeding the magnetic
energy. From equation \ref{l_kin_ep} we see that the ratio between
proton energy and electron energy is $m_{\rm p}/(\gamma_{\rm e,min}
m_{\rm e})$, which is larger than unity in all cases of minimum
electron energy discussed in Section 2. Thus, the only energy source
for accelerating electrons should be the kinetic energy of protons
and such acceleration process would occur in shock process. The
electrons are preheated up to average energy of protons heated in
the shocks, and then the energy is converted from protons to
electrons in the diffusive shock acceleration process
\citep*[e.g.,][]{2006ApJ...653..325A,2009ApJ...690..244A,2011ApJ...726...75S}.
Several authors have investigated these scenarios, both analytically
and in particle-in-cell simulations
\citep{2006ApJ...653..325A,2009ApJ...690..244A,2011ApJ...726...75S}.
From this point of view, the electron-proton content is preferred
for this sample of radio loud quasars, or at least the
electron-proton dominates dynamically over the electron-position pair content.
As discussed that in Section 2, only the normalization is changed if
the different plasma composition is considered, which means that
most of the statistic results derived in this work is independent of
the jet composition.

No correlation is found between the Eddington ratio and the Lorentz
factor of the jet, which implies that the jet acceleration may not
be related to the properties of the accretion disk. The results
imply that the BZ mechanism may dominate over the BP mechanism for
jet acceleration at least in radio-loud. \citet{2000ApJ...543L.111L}
found that the black holes in RL AGNs are systematically heavier
than those in radio-quiet counterparts, which may imply that heavy
black holes are spinning rapidly, and therefore the jets can be
easily accelerated by the field lines threading the horizons of
these black holes. This is roughly consistent with the conclusion of
this work.

The origin of the chaotic magnetic fields in the jets is still
unclear. The jets are accelerated by the magnetic fields near the
black hole horizon or those threading the rotational accretion disk.
It is therefore reasonable to postulate that the strength of the
field in the jets is related to the field driving the jets in some
way. No significant correlation is present between the magnetic
field strength at $10R_{\rm S}$ and the bulk Lorentz factor of the
jet components $\Gamma$. The independence of the bulk velocity of
the jets on the magnetic field strength implies that the Lorentz
factor of the jet components is mainly governed by the black hole
spin.




\section*{Acknowledgments}
We thank the referee for his/her helpful comments. This work is
supported by the National Basic Research Program of China (grant
2009CB824800), the NSFC (grants 11173043, 11073039, 11121062 and
10833002), and the CAS/SAFEA International Partnership Program for
Creative Research Teams (KJCX2-YW-T23).

\clearpage

\begin{deluxetable}{ccccccclc}
\tablewidth{0pt}
 \tablecaption{Data for the sample.\label{tbl-2}} \tablehead{
\colhead{Source}&\colhead{Class}&\colhead{z}&\colhead{log $\Gamma$}&
\colhead{$n_{1}$} &\colhead{$B_{1}$}&log $M_{\textrm{BH}}$ & Refs. &log
$L_{\textrm{BLR}}$
\\& & & & ($\textrm{cm}^{-3}$) &(gauss)&($M_{\odot}$)& &(erg
$\textrm{s}^{-1}$)\\(1)&(2)&(3)&(4)&(5)&(6)&(7)&(8)&(9)} \startdata

0007+106 & Qc & 0.089 & 0.279 & 1.32E+04& 0.030 &  8.29 &K07 & 44.14\\
0016+731 & Qc & 1.781 & 1.248 & 6.75E+04&0.546 & 8.93 & Z09 & 44.98\\
0035+413 & Qc & 1.353 & 1.283 & 5.01E+06&0.724 & 8.53 & Z09 & 44.64\\
0106+013 & Qc & 2.107 & 1.378 & 1.15E+05&0.687 & 8.83 & Z09 & 46.15\\
0112$-$017 & Qc & 1.365 & 0.230 & 9.52E+04& 0.089 & 7.85 & Z09 & 45.26\\
0133+207 & Ql & 0.425 & 1.456 & 2.09E+06& 0.230 & 9.45 & L06 & 45.02\\
0133+476 & Qc & 0.859 & 0.568 & 4.42E+04& 0.254 & 8.30 & Z09 & 44.47\\
0153+744 & Qc & 2.338 & 1.090 & 2.23E+06& 0.227 & 9.82 & S03 & 46.14\\
0202+149 & Qc & 0.405 & 0.820 & 1.63E+04& 0.240 & 9.60 & L03 & \nodata\\
0208$-$512 & Qc & 1.003 & 1.505 & 2.07E+04& 0.753 & 9.21 & F04 & 45.19\\
0212+735 & Qc & 2.367 & 0.924 & 5.06E+05& 0.498 & 6.96 & L06 & 44.95\\
0219+428 & BL & 0.444 & 1.580 & 1.89E+04& 0.201 & 8.60 & L03 & \nodata\\
0235+164 & BL & 0.940 & 1.823 & 1.37E+05& 2.125 & 10.22 & W04 & 43.86\\
0316+413 & G  & 0.017 & 0.114 & 4.76E+03& 0.012 & 8.51 & P06 & 42.70\\
0333+321 & Qc & 1.263 & 1.439 & 1.01E+05& 0.543 & 9.25 & L06 & 45.93\\
0336$-$019 & Qc & 0.852 & 1.294 & 9.73E+03& 0.404 & 8.98 & W02 & 45.00\\
0420$-$014 & Qc & 0.915 & 1.193 & 1.01E+05& 0.466 & 8.41 & L06 & 44.92\\
0430+052 & G  & 0.033 & 0.778 & 3.43E+02& 0.019 & 7.74 & P04 & 42.93\\
0440$-$003 & Qc & 0.844 & 1.462 & 2.89E+05& 0.526 & 8.81 & F04 & 44.77\\
0458$-$020 & Qc & 2.286 & 1.140 & 4.42E+04& 0.376 & 8.66 & F04 & 45.32\\
0528+134 & Qc & 2.060 & 1.588 & 3.32E+05& 1.187 & 10.20 & L03 & \nodata\\
0605$-$085 & Qc & 0.872 & 1.127 & 4.77E+05& 0.579 & 8.43 & Z09 & 44.62\\
0607$-$157 & Qc & 0.324 & 0.591 & 1.39E+04& 0.196 & 7.32 & L06 & 43.56\\
0716+714 & BL & 0.300 & 1.695 & 6.91E+04& 0.258 & 8.10 & L03 & \nodata\\
0723+679 & Ql & 0.846 & 1.185 & 3.82E+05& 0.420 & 8.67 & W02 & 44.80\\
0735+178 & BL & 0.424 & 1.401 & 4.70E+04& 0.353 & 8.40 & C03 & \nodata\\
0736+017 & Qc & 0.191 & 1.090 & 1.05E+03& 0.096 & 8.47 & M01 & 44.18\\
0738+313 & Qc & 0.630 & 0.431 & 2.64E+04& 0.126 & 9.40 & W02 & 45.78\\
0745+241 & Qc & 0.409 & 0.934 & 1.20E+04& 0.166 & 7.92 & S08 & \nodata\\
0748+126 & Qc & 0.889 & 1.121 & 2.69E+04& 0.394 & 8.15 & L06 & 44.95\\
0754+100 & BL & 0.266 & 1.107 & 8.83E+03& 0.145 & 10.14 & C03 & \nodata\\
0804+499 & Qc & 1.432 & 0.944 & 6.67E+04& 0.279 & 9.39 & L06 & 45.39\\
0823+033 & BL & 0.506 & 1.164 & 8.33E+03& 0.183 & 8.83 & C03 & 43.40\\
0827+243 & Qc & 0.939 & 1.507 & 1.23E+05& 0.568 & 9.80 & L03 & 44.93\\
0829+046 & BL & 0.180 & 1.246 & 1.97E+03& 0.111 & 8.46 & W05 & \nodata\\
0836+710 & Qc & 2.172 & 1.467 & 1.50E+06& 0.607 & 9.36 & L06 & 46.43\\
0850+581 & Qc & 1.322 & 1.575 & 2.02E+06& 0.275 & 8.49 & L06 & 45.66\\
0851+202 & BL & 0.306 & 1.155 & 1.87E+04& 0.147 & 8.79 & C03 & 43.60\\
0859$-$140 & Ql & 1.339 & 1.225 & 4.13E+04& 0.379 & 8.87 & Z09 & 45.74\\
0906+015 & Qc & 1.018 & 1.086 & 7.05E+04& 0.461 & 8.55 & L06 & 45.11\\
0906+430 & Qc & 0.670 & 1.041 & 2.95E+03& 0.191 & 7.90 & W02 & 43.34\\
0917+449 & Qc & 2.180 & 1.117 & 8.70E+04& 0.297 & 9.31 & S08 & 45.21\\
0923+392 & Qc & 0.695 & 1.637 & 1.77E+07& 0.801 & 9.28 & W02 & 45.79\\
0945+408 & Qc & 1.252 & 1.356 & 1.55E+04& 0.388 & 8.60 & L06 & 45.59\\
0953+254 & Qc & 0.712 & 1.350 & 1.39E+05& 0.373 & 9.00 & W02 & 44.97\\
0954+658 & BL & 0.368 & 1.013 & 4.17E+03& 0.115 & 8.53 & F04 & 42.63\\
1012+232 & Qc & 0.565 & 0.968 & 5.70E+03& 0.197 & 8.69 & Z09 & 45.16\\
1015+359 & Qc & 1.226 & 1.182 & 1.07E+05& 0.461 & 9.11 & S08 & 45.98\\
1040+123 & Qc & 1.029 & 1.072 & 1.10E+06& 0.461 & 8.76 & L06 & 45.11\\
1055+018 & Qc & 0.888 & 0.740 & 2.80E+04& 0.338 & 8.45 & Z09 & 44.53\\
1101+384 & BL & 0.031 & 0.519 & 1.67E+03& 0.008 & 8.22 & W05 & 41.40\\
1127$-$145 & Qc & 1.187 & 1.843 & 9.81E+05& 1.161 & 9.18 & Z09 & 45.77\\
1128+385 & Qc & 1.733 & 0.322 & 7.96E+04& 0.169 & 9.29 & S08 & 46.26\\
1137+660 & Ql & 0.646 & 0.708 & 6.98E+05& 0.110 & 9.31 & L06 & 45.85\\
1156+295 & Qc & 0.729 & 1.346 & 1.05E+04& 1.107 & 8.54 & L06 & 44.90\\
1219+285 & BL & 0.102 & 0.845 & 1.12E+04& 0.054 & 7.40 & L03 & 42.25\\
1222+216 & Ql & 0.435 & 1.438 & 1.21E+04& 0.253 & 8.44 & F04 & 44.73\\
1226+023 & Qc & 0.158 & 1.436 & 1.29E+05& 0.274 & 8.95 & P04 & 45.59\\
1253$-$055 & Qc & 0.538 & 1.017 & 3.83E+04& 0.391 & 8.28 & L06 & 44.64\\
1302$-$102 & Qc & 0.286 & 0.799 & 1.33E+04& 0.075 & 8.58 & K08 & 44.91\\
1308+326 & BL & 0.996 & 1.656 & 4.41E+04& 0.934 & 9.24 & W04 & 45.12\\
1334$-$127 & Qc & 0.539 & 0.991 & 1.06E+04& 0.373 & 7.98 & L06 & 44.18\\
1345+125 & G  & 0.121 & 0.806 & 5.13E+05& 0.235 & 7.80 & W09 & \nodata\\
1406$-$076 & Q  & 1.494 & 1.786 & 1.01E+05& 0.948 & 9.40 & L03 & 45.47\\
1458+718 & Qc & 0.905 & 1.117 & 4.58E+04& 0.271 & 8.77 & L06 & 45.47\\
1508$-$055 & Ql & 1.191 & 1.639 & 8.48E+04& 0.566 & 8.97 & Z09 & 45.52\\
1510$-$089 & Qc & 0.360 & 1.446 & 4.26E+04& 0.345 & 8.65 & W02 & 44.65\\
1532+016 & Qc & 1.420 & 1.696 & 5.15E+06& 1.074 & 8.73 & Z09 & 44.84\\
1546+027 & Qc & 0.412 & 0.380 & 1.85E+04& 0.095 & 8.31 & O02 & 44.68\\
1606+106 & Qc & 1.226 & 1.631 & 1.57E+05& 1.106 & 9.50 & L03 & \nodata\\
1611+343 & Qc & 1.401 & 1.919 & 1.07E+06& 1.603 & 9.60 & L06 & 45.91\\
1618+177 & Ql & 0.555 & 1.013 & 1.17E+06& 0.158 & 9.65 & L06 & 46.13\\
1622$-$297 & Q  & 0.815 & 1.328 & 6.06E+04& 0.669 & 9.10 & L03 & \nodata\\
1633+382 & Qc & 1.814 & 1.196 & 9.90E+04& 0.443 & 9.67 & L06 & 45.84\\
1637+826 & G  & 0.023 & 0.041 & 4.94E+02& 0.007 & 8.78 & G09 & \nodata\\
1641+399 & Qc & 0.593 & 1.470 & 5.91E+05& 1.043 & 9.42 & W02 & 45.47\\
1642+690 & Qc & 0.751 & 1.223 & 6.14E+03& 0.265 & 7.76 & W02 & 43.86\\
1652+398 & BL & 0.034 & 0.623 & 1.24E+04& 0.019 & 8.62 & W05 & 41.36\\
1655+077 & Qc & 0.621 & 1.442 & 2.28E+05& 0.698 & 7.28 & L06 & 43.62\\
1656+053 & Qc & 0.879 & 0.898 & 1.12E+06& 0.297 & 9.62 & W02 & 46.26\\
1721+343 & Ql & 0.205 & 1.161 & 7.05E+05& 0.101 & 8.04 & W02 & 44.63\\
1730$-$130 & Qc & 0.902 & 1.418 & 1.32E+05& 0.756 & 9.30 & L03 & 44.83\\
1749+096 & BL & 0.320 & 0.949 & 2.62E+03& 0.271 & 8.34 & Z09 & \nodata\\
1749+701 & BL & 0.770 & 1.072 & 2.29E+04& 0.208 & 10.39 & C03 & \nodata\\
1803+784 & BL & 0.684 & 0.041 & 1.04E+05& 0.079 & 7.92 & L06 & 44.56\\
1807+698 & BL & 0.051 & 1.021 & 6.15E+02& 0.054 & 8.51 & B03 & 41.40\\
1823+568 & BL & 0.664 & 0.672 & 2.95E+04& 0.200 & 9.26 & Wu02 & 43.32\\
1828+487 & Ql & 0.692 & 1.176 & 1.30E+04& 0.237 & 9.85 & L06 & 45.25\\
1845+797 & Qc & 0.057 & 0.580 & 1.02E+04& 0.018 & 8.46 & P04 & 42.97\\
1921$-$293 & Qc & 0.352 & 0.954 & 3.90E+05& 0.146 & 8.38 & Z09 & 43.67\\
1928+738 & Qc & 0.302 & 1.061 & 6.88E+03& 0.181 & 8.91 & W02 & 45.18\\
2131$-$021 & BL & 1.285 & 0.892 & 7.38E+04& 0.365 & 10.21 & C03 & 43.66\\
2134+004 & Qp & 1.932 & 0.322 & 3.41E+05& 0.195 & 8.50 & Z09 & 46.29\\
2145+067 & Qc & 0.999 & 0.556 & 8.31E+04& 0.335 & 8.87 & L06 & 45.78\\
2200+420 & BL & 0.069 & 0.845 & 6.27E+02& 0.049 & 8.23 & W02 & 42.38\\
2201+315 & Qc & 0.298 & 0.851 & 8.23E+03& 0.095 & 8.43 & W02 & 45.46\\
2223-052 & Qc & 1.404 & 1.520 & 4.53E+04& 0.606 & 8.54 & Z09 & 45.62\\
2230+114 & Qc & 1.037 & 1.446 & 5.51E+04& 0.476 & 8.64 & Z09 & 45.89\\
2243$-$123 & Qc & 0.630 & 1.100 & 7.80E+04& 0.352 & 8.32 & P05 & 45.28\\
2251+158 & Qc & 0.859 & 1.511 & 1.67E+05& 0.493 & 9.17 & W02 & 45.68\\
2345$-$167 & Qc & 0.576 & 0.204 & 3.55E+04& 0.091 & 8.47 & L06 & 44.38\\
\enddata
\tablecomments{Column (1): IAU name; Column (2): classification of
the source (Q$=$quasars; Qc$=$core-dominated quasars;
Ql$=$lobe-dominated quasars; Qp$=$GHz peaked quasars; BL$=$BL Lac
objects; G$=$radio galaxies). Column (3): redshift. Column (4): the
jet Lorentz factor $\Gamma$. Column(5): density of electrons $n_{1}$.
Column(6): magnetic field strength. Column (7): black hole mass.
Column (8): references for black hole mass.
Column (9): the total luminosity in broad emission lines
$L_{\textrm{BLR}}$.} \tablerefs{ B03: \citet{2003ApJ...583..134B};
C03: \citet{2003ApJ...599..147C}; F04: \citet{2004ApJ...602..103F};
G09: \citet{2009ApJ...706..404G}; K07: \citet{2007ApJ...661..660K};
K08: \citet{2008ApJ...687..767K};
 L03: \citet{2003MNRAS.340..632L}; L06: \citet{2006ApJ...637..669L}; M01:
\citet{2001MNRAS.327..199M}; O02: \citet{2002ApJ...576...81O}; P04:
\citet{2004ApJ...613..682P}; P05: \citet{2005MNRAS.361..919P}; P06:
\citet {2006A&A...455..173P}; S03: \citet{2003ApJ...583..124S};
S08:\citet{2008ApJ...680..169S}; W02: \citet{2002ApJ...579..530W};
W05: \citet{2005ApJ...631..762W}; W09: \citet{2009MNRAS.398.1905W};
Wu02: \citet{2002A&A...389..742W}; Z09:
\citet{2009RAA.....9..293Z};}
\end{deluxetable}

\begin{deluxetable}{cccccccc}
\tablewidth{0pt} \tablecaption{Spearman Partial Rank Correlation
Analysis of the Sample}
\tablehead{\colhead{Sample}&\colhead{N}&\colhead{Correlation:A,B}&Variable:C&\colhead{$r_{\textrm{AB}}$}&\colhead{significance}&\colhead{$r_{\textrm{AB,C}}$}&\colhead{significance}\\
&&&&&level(\%)&&\\ (1)&(2)&(3)&(4)&(5)&(6)&(7)&(8)} \startdata All &
101 &
$M_{\textrm{BH}}$,$\Gamma$ &z&0.357&99.98&0.240&2.41\\
 & & $\Gamma$,z &$M_{\textrm{BH}}$& 0.385&99.99& \nodata& \nodata\\
 & & $M_{\textrm{BH}}$,z&$\Gamma$&0.400&99.99& \nodata& \nodata\\
 Quasars&77&$M_{\textrm{BH}}$,$\Gamma$&z&0.376&99.92&0.305&2.70\\
 & &$\Gamma$,z&$M_{\textrm{BH}}$&0.311&99.41& \nodata& \nodata\\
 & &$M_{\textrm{BH}}$,z&$\Gamma$&0.327&99.63& \nodata& \nodata\\
with $L_{\textrm{BLR}}$ & 87 &
$L_{\textrm{bol}}/L_{\textrm{Edd}}$,$\Gamma$ &z&
0.099 &63.94& \nodata& \nodata\\
sources within &23&$M_{\textrm{BH}}$,$\Gamma$&z&0.590&99.64&0.580&2.89\\
$0.8<z<1.2$& &$\Gamma$,z&$M_{\textrm{BH}}$&0.243&73.54& \nodata& \nodata\\
 & &$M_{\textrm{BH}}$,z&$\Gamma$&0.132&45.01& \nodata& \nodata\\

\enddata
\tablecomments{ Here $r_{\textrm{AB}}$ is the rank correlation
coefficient of the two variables, and $r_{\textrm{AB,C}}$ is the
partial rank correlation coefficient. Column (6) is the significance
level of the rank correlation. Column (8) is the significance of the
partial rank correlation equivalent to the deviation from a unit
variance normal distribution if there is no correlation present. }
\end{deluxetable}

\clearpage
\begin{figure}
\epsscale{.80} \plotone{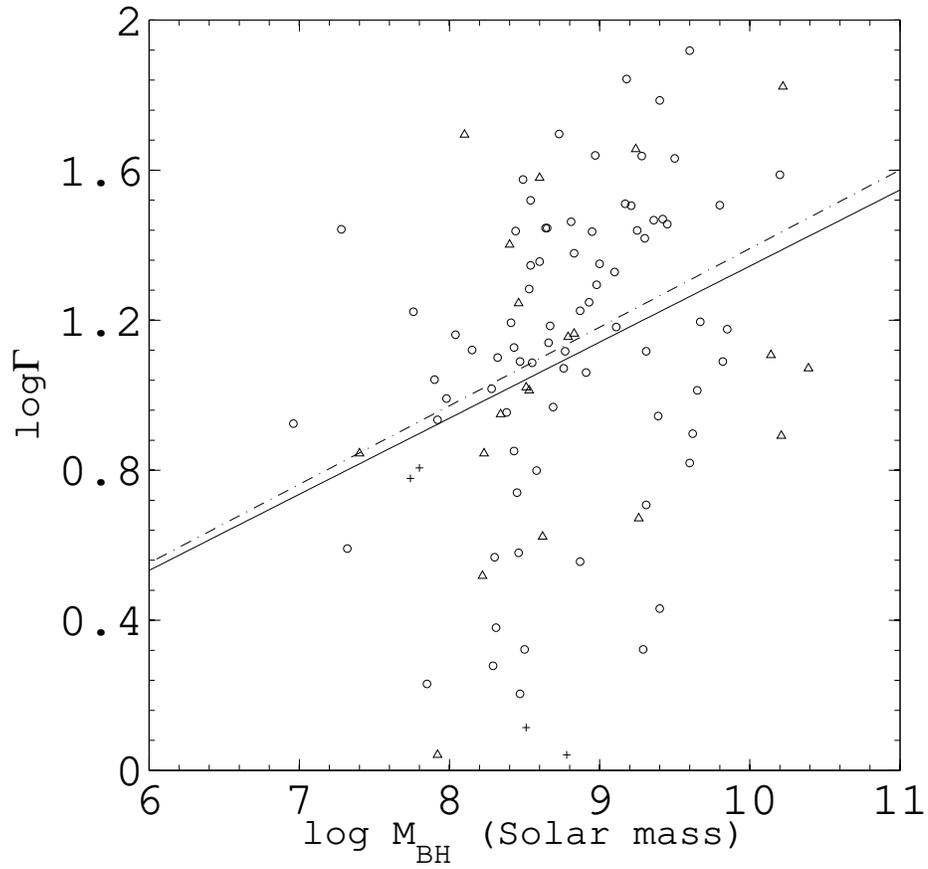} \caption{The relation between black
hole mass and the bulk Lorentz factor of the jet components. The
circles represent quasars, and the triangle represents BL Lac
objects, while the crosses represent radio galaxies. The solid line
is the fitted line for the whole sample using the least square
method while the dot-dashed line is fitted for quasars only.
\label{mbh_gamma}}
\end{figure}

\begin{figure}
\epsscale{.80} \plotone{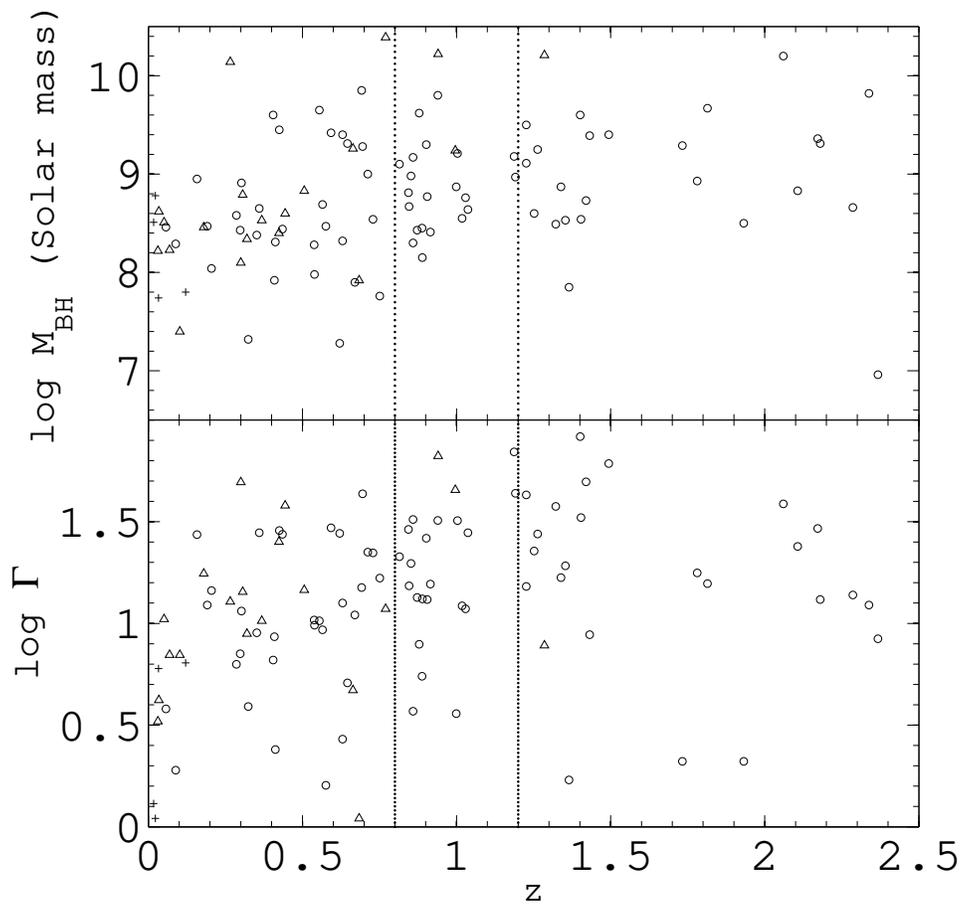} \caption{The relation between black
hole mass and redshift $z$ (the upper panel). The lower panel is the
relation between the bulk Lorentz factor of the jet components and
redshift $z$. The two vertical dotted lines correspond to $z=0.8$
and $1.2$, respectively.  \label{mbh_gamma_z}}
\end{figure}

\begin{figure}
\epsscale{.80} \plotone{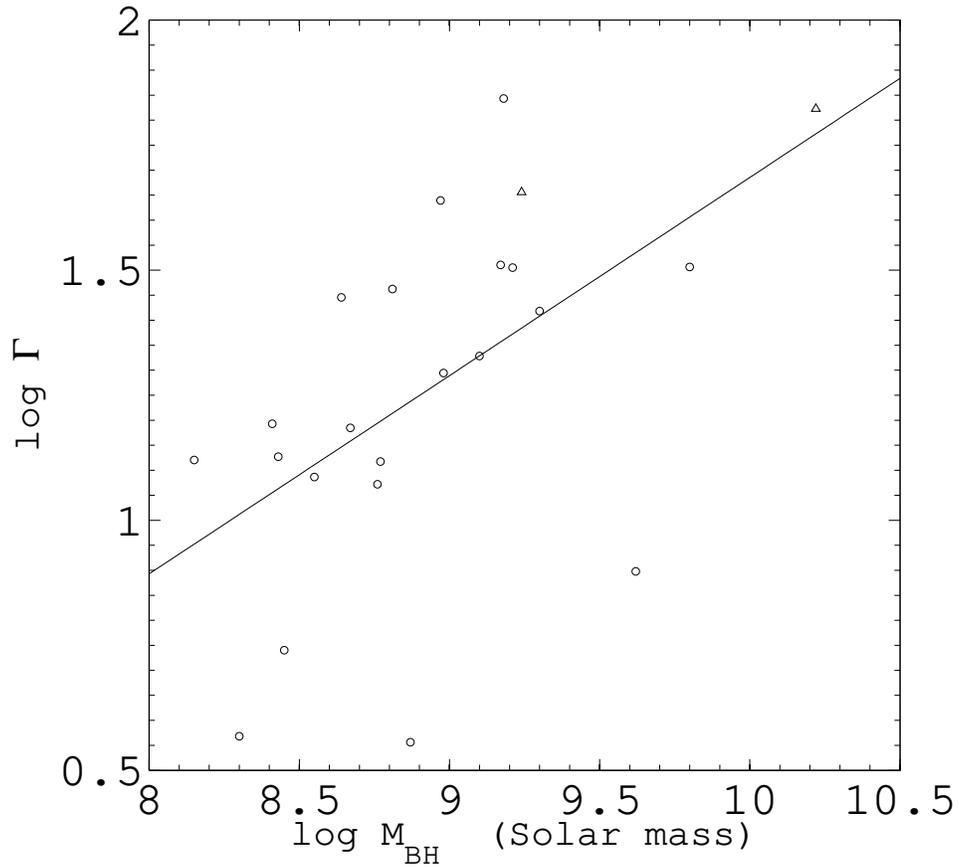} \caption{The relation between black
hole mass and the bulk Lorentz factor of the jet components for the
sub-sample of the sources with redshift $0.8<z<1.2$. The symbols are
the same as Figure \ref{mbh_gamma}. \label{mbh_gamma_sub}}
\end{figure}

\begin{figure}
\epsscale{.80} \plotone{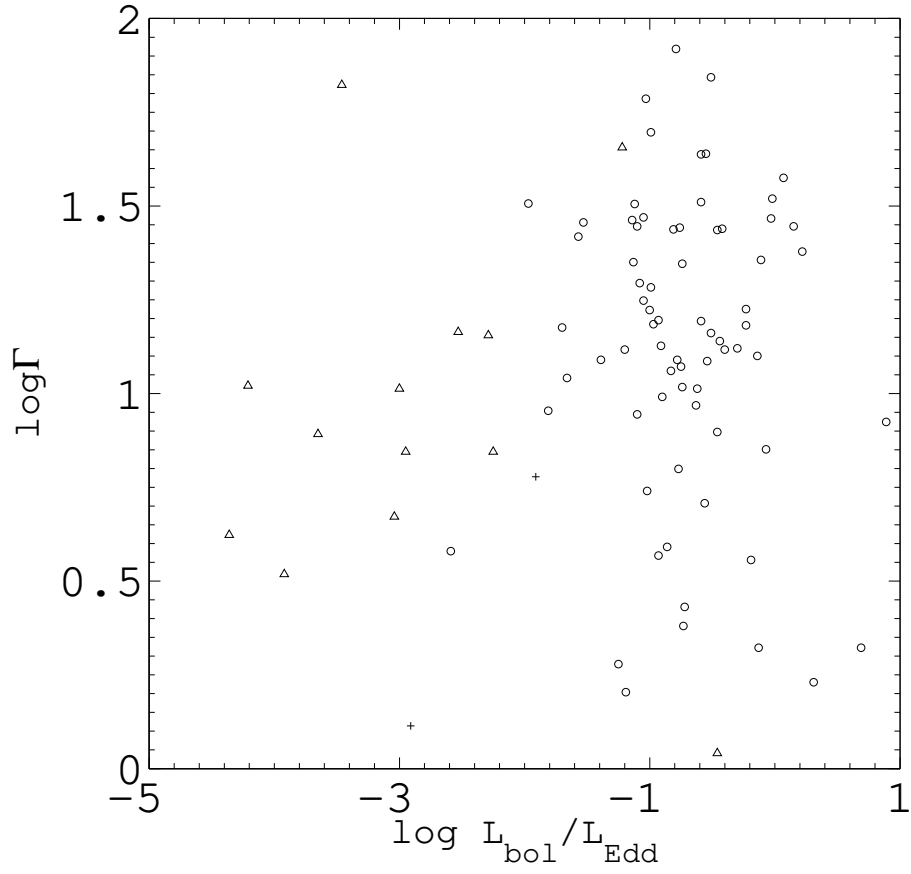} \caption{The relation between the
Eddington ratio and the bulk Lorentz factor of jet. The symbols are
the same as Figure \ref{mbh_gamma}. \label{eddratio_gamma}}
\end{figure}

\begin{figure}
\epsscale{.80} \plotone{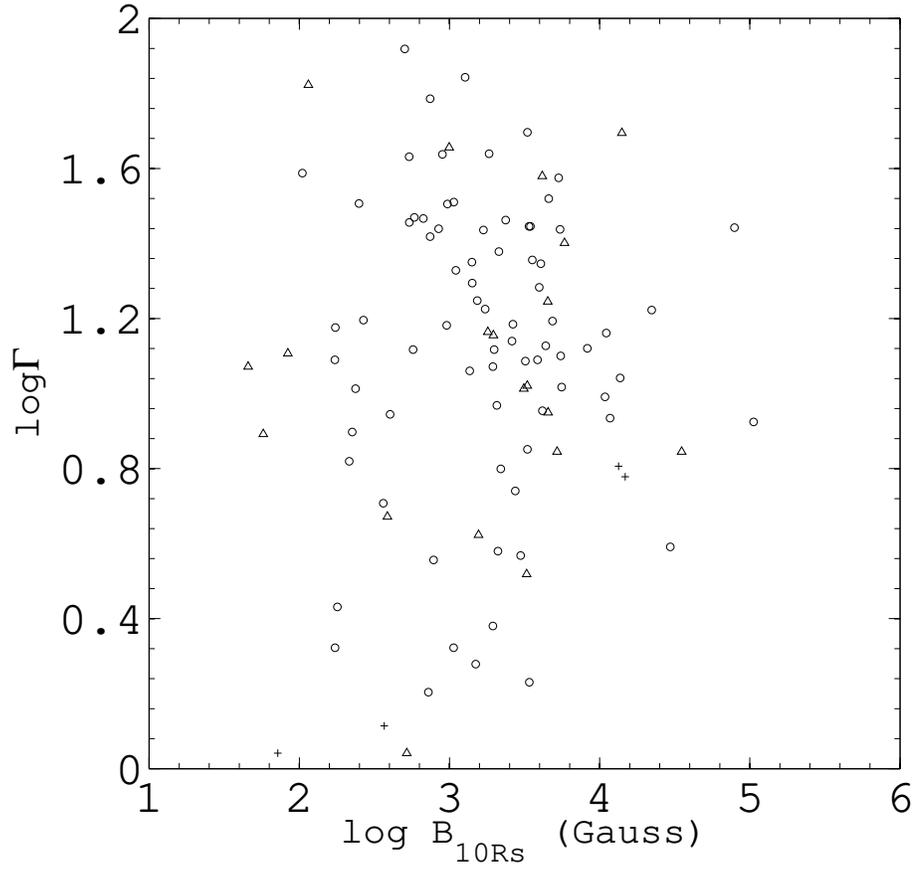} \caption{The relation between
magnetic field strength at $10R_{\rm S}$ and the bulk Lorentz factor
of jet. The symbols are the same as Figure \ref{mbh_gamma}.
\label{b_gamma}}
\end{figure}

\begin{figure}
\epsscale{.80} \plotone{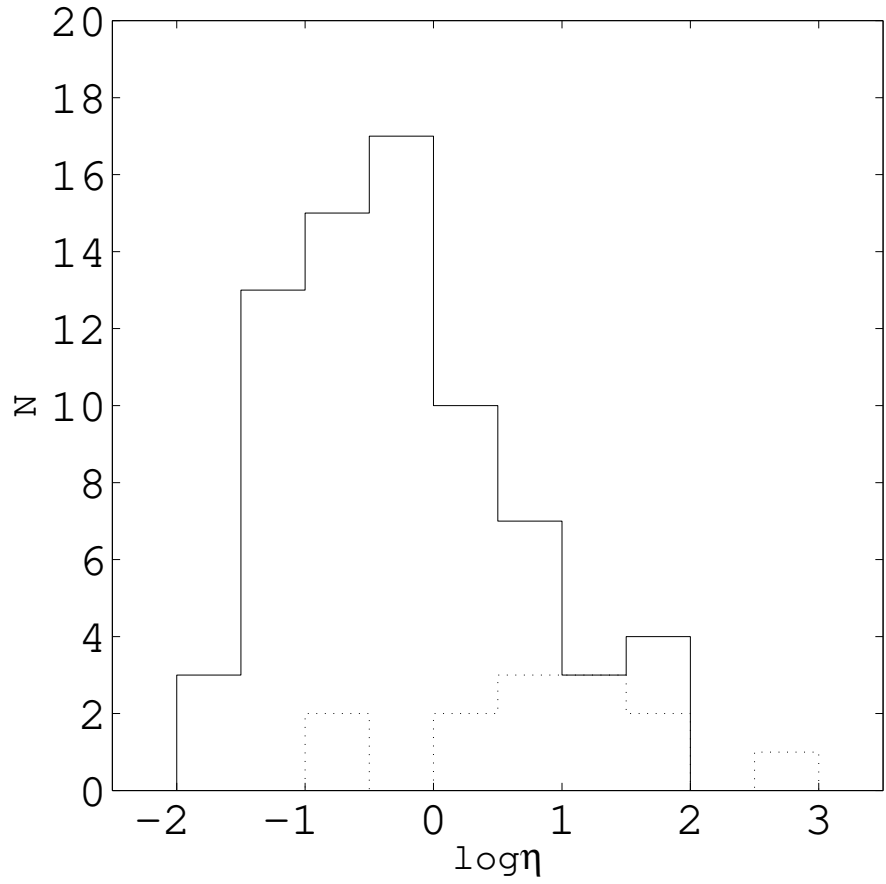} \caption{The distribution of jet
efficiency $\eta_{\rm jet}$ (solid line: quasars, and dotted line:
BL Lac objects).  \label{eta number}}
\end{figure}

\begin{figure}
\epsscale{.80} \plotone{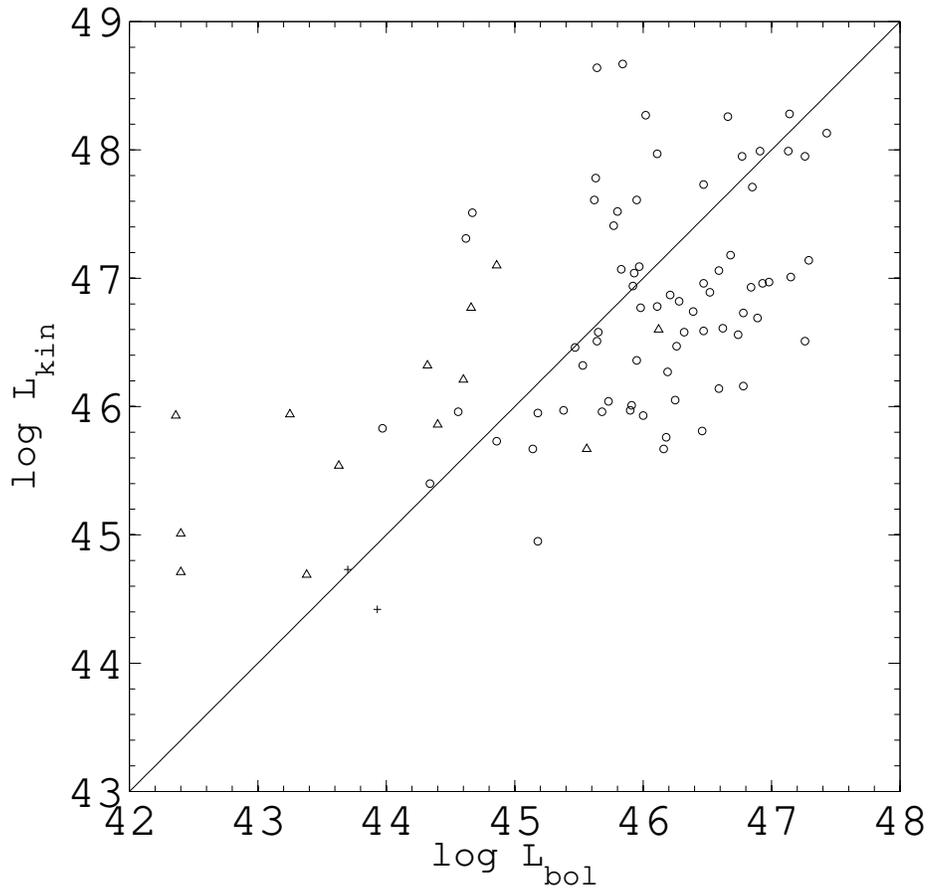} \caption{The relation between
bolometric luminosity and kinetic power of jets. The line
correspond to jet efficiency of $\eta$=1.
The symbols are the same as Figure \ref{mbh_gamma}.\label{jet efficiency}}
\end{figure}

\begin{figure}
\epsscale{.80} \plotone{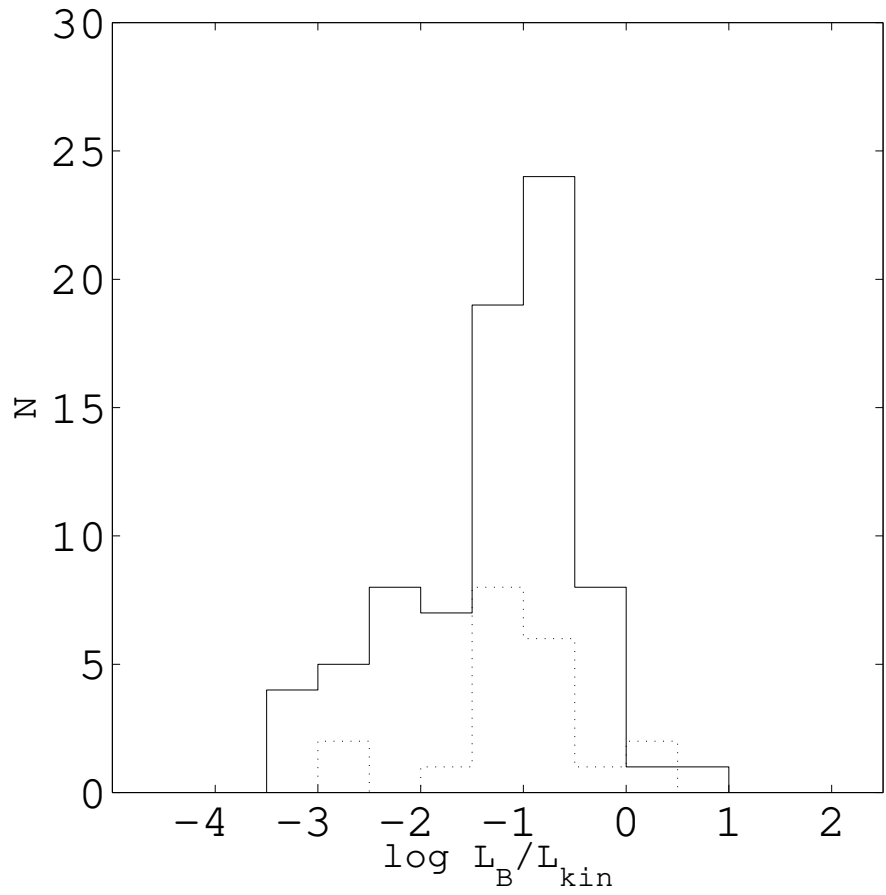} \caption{The distribution of radio between magnetic energy flux and bulk kinetic power
 (solid line: quasars, and dotted line:
BL Lac objects).  \label{l_kin l_b}}
\end{figure}


\begin{thebibliography}{}

\bibitem[Amato
\& Arons(2006)]{2006ApJ...653..325A} Amato, E., \& Arons, J.\ 2006, \apj, 653, 325

\bibitem[Amano
\& Hoshino(2009)]{2009ApJ...690..244A} Amano, T., \& Hoshino, M.\ 2009, \apj, 690, 244


\bibitem[Barth et al.(2003)]{2003ApJ...583..134B} Barth, A.~J., Ho, L.~C.,
\& Sargent, W.~L.~W.\ 2003, \apj, 583, 134

\bibitem[Blandford \&
K\"{o}nigl(1979)]{1979ApJ...232...34B} Blandford, R.~D., K\"{o}nigl,
A.\ 1979, \apj, 232, 34


\bibitem[Blandford
\& Payne(1982)]{1982MNRAS.199..883B} Blandford, R.~D., \& Payne,
D.~G.\ 1982, \mnras, 199, 883

\bibitem[Blandford
\& Znajek(1977)]{1977MNRAS.179..433B} Blandford, R.~D., \& Znajek,
R.~L.\ 1977, \mnras, 179, 433

\bibitem[Blundell et al.(2006)]{2006ApJ...644L..13B} Blundell, K.~M.,
Fabian, A.~C., Crawford, C.~S., Erlund, M.~C., \& Celotti, A.\ 2006,
\apjl, 644, L13

\bibitem[Cao(2003)]{2003ApJ...599..147C} Cao, X.\ 2003, \apj, 599, 147


\bibitem[Cao(2011)]{2011ApJ...737...94C} Cao, X.\ 2011, \apj, 737, 94

\bibitem[Cao
\& Jiang(1999)]{1999MNRAS.307..802C} Cao, X., \& Jiang, D.~R.\ 1999,
\mnras, 307, 802

\bibitem[Cao
\& Jiang(2001)]{2001MNRAS.320..347C} Cao, X., \& Jiang, D.~R.\ 2001,
\mnras, 320, 347

\bibitem[Cao
\& Li(2008)]{2008MNRAS.390..561C} Cao, X., \& Li, F.\ 2008, \mnras,
390, 561


\bibitem[Celotti
\& Fabian(1993)]{1993MNRAS.264..228C} Celotti, A., \& Fabian, A.~C.\ 1993, \mnras, 264, 228


\bibitem[Celotti et al.(1997)]{1997MNRAS.286..415C} Celotti, A., Padovani,
P., \& Ghisellini, G.\ 1997, \mnras, 286, 415







\bibitem[Fan
\& Cao(2004)]{2004ApJ...602..103F} Fan, Z.-H., \& Cao, X.\ 2004,
\apj, 602, 103

\bibitem[Ghisellini et al.(1993)]{1993ApJ...407...65G} Ghisellini, G.,
Padovani, P., Celotti, A., \& Maraschi, L.\ 1993, \apj, 407, 65

\bibitem[Ghosh
\& Abramowicz(1997)]{1997MNRAS.292..887G} Ghosh, P., \& Abramowicz,
M.~A.\ 1997, \mnras, 292, 887

\bibitem[Godfrey et al.(2009)]{2009ApJ...695..707G} Godfrey, L.~E.~H.,
Bicknell, G.~V., Lovell, J.~E.~J., et al.\ 2009, \apj, 695, 707

\bibitem[Gu et al.(2009)]{2009MNRAS.396..984G} Gu, M., Cao, X.,
\& Jiang, D.~R.\ 2009, \mnras, 396, 984

\bibitem[Guijosa
\& Daly(1996)]{1996ApJ...461..600G} Guijosa, A., \& Daly, R.~A.\
1996, \apj, 461, 600

\bibitem[G{\"u}ltekin et al.(2009)]{2009ApJ...706..404G} G{\"u}ltekin, K.,
Cackett, E.~M., Miller, J.~M., et al.\ 2009, \apj, 706, 404





\bibitem[Hirotani(2005)]{2005ApJ...619...73H} Hirotani, K.\ 2005, \apj,
619, 73

\bibitem[Hughes
\& Blandford(2003)]{2003ApJ...585L.101H} Hughes, S.~A., \&
Blandford, R.~D.\ 2003, \apjl, 585, L101


\bibitem[Jiang et al.(1998)]{1998ApJ...494..139J} Jiang, D.~R., Cao, X.,
\& Hong, X.\ 1998, \apj, 494, 139


\bibitem[Kawakatu et al.(2007)]{2007ApJ...661..660K} Kawakatu, N.,
Imanishi, M., \& Nagao, T.\ 2007, \apj, 661, 660

\bibitem[Kellermann et al.(2004)]{2004ApJ...609..539K} Kellermann, K.~I.,
Lister, M.~L., Homan, D.~C., et al.\ 2004, \apj, 609, 539

\bibitem[Kellermann et al.(1989)]{1989AJ.....98.1195K} Kellermann, K.~I.,
Sramek, R., Schmidt, M., Shaffer, D.~B., \& Green, R.\ 1989, \aj,
98, 1195

\bibitem[Kim et al.(2008)]{2008ApJ...687..767K} Kim, M., Ho, L.~C., Peng,
C.~Y., et al.\ 2008, \apj, 687, 767


\bibitem[Kino et al.(2012)]{2012ApJ...751..101K} Kino, M., Kawakatu, N.,
\& Takahara, F.\ 2012, \apj, 751, 101

\bibitem[Kino
\& Takahara(2004)]{2004MNRAS.349..336K} Kino, M., \& Takahara, F.\
2004, \mnras, 349, 336


\bibitem[K{\"o}nigl(1981)]{1981ApJ...243..700K} K{\"o}nigl, A.\ 1981, \apj, 243,
700



\bibitem[L{\"a}hteenm{\"a}ki
\& Valtaoja(1999)]{1999ApJ...521..493L} L{\"a}hteenm{\"a}ki, A., \&
Valtaoja, E.\ 1999, \apj, 521, 493

\bibitem[Laor(2000)]{2000ApJ...543L.111L} Laor, A.\ 2000, \apjl, 543, L111




\bibitem[Li
\& Cao(2012)]{2012ApJ...753...24L} Li, S.-L., \& Cao, X.\ 2012,
\apj, 753, 24

\bibitem[Liang
\& Liu(2003)]{2003MNRAS.340..632L} Liang, E.~W., \& Liu, H.~T.\
2003, \mnras, 340, 632

\bibitem[Lister
\& Homan(2005)]{2005AJ....130.1389L} Lister, M.~L., \& Homan, D.~C.\
2005, \aj, 130, 1389

\bibitem[Liu et al.(2006)]{2006ApJ...637..669L} Liu, Y., Jiang, D.~R.,
\& Gu, M.~F.\ 2006, \apj, 637, 669

\bibitem[Livio et al.(1999)]{1999ApJ...512..100L} Livio, M., Ogilvie,
G.~I., \& Pringle, J.~E.\ 1999, \apj, 512, 100

\bibitem[Marscher(1987)]{1987slrs.work..280M} Marscher, A.~P.\ 1987,
Superluminal Radio Sources, 280




\bibitem[McKinney et al.(2012)]{2012arXiv1201.4163M} McKinney, J.~C.,
Tchekhovskoy, A., \& Blandford, R.~D.\ 2012, \mnras, 423, 3083





\bibitem[McLure
\& Dunlop(2001)]{2001MNRAS.327..199M} McLure, R.~J., \& Dunlop,
J.~S.\ 2001, \mnras, 327, 199



\bibitem[Moderski
\& Sikora(1996)]{1996MNRAS.283..854M} Moderski, R., \& Sikora, M.\
1996, \mnras, 283, 854


\bibitem[Netzer(1990)]{1990agn..conf...57N} Netzer, H.\ 1990, Active
Galactic Nuclei, 57

\bibitem[Oshlack et al.(2002)]{2002ApJ...576...81O} Oshlack, A.~Y.~K.~N.,
Webster, R.~L., \& Whiting, M.~T.\ 2002, \apj, 576, 81



\bibitem[Panessa et
al.(2006)]{2006A&A...455..173P} Panessa, F., Bassani, L., Cappi, M.,
et al.\ 2006, \aap, 455, 173

\bibitem[Park
\& Blackman(2010)]{2010MNRAS.403.1993P} Park, K., \& Blackman,
E.~G.\ 2010, \mnras, 403, 1993


\bibitem[Peterson et al.(2004)]{2004ApJ...613..682P} Peterson, B.~M.,
Ferrarese, L., Gilbert, K.~M., et al.\ 2004, \apj, 613, 682

\bibitem[Pian et al.(2005)]{2005MNRAS.361..919P} Pian, E., Falomo, R.,
\& Treves, A.\ 2005, \mnras, 361, 919




\bibitem[Rawlings et al.(1989)]{1989MNRAS.240..701R} Rawlings, S.,
Saunders, R., Eales, S.~A., \& Mackay, C.~D.\ 1989, \mnras, 240, 701

\bibitem[Readhead(1994)]{1994ApJ...426...51R} Readhead, A.~C.~S.\ 1994,
\apj, 426, 51

\bibitem[Reynolds et al.(1996)]{1996MNRAS.283..873R} Reynolds, C.~S.,
Fabian, A.~C., Celotti, A., \& Rees, M.~J.\ 1996, \mnras, 283, 873

\bibitem[Shen et al.(2008)]{2008ApJ...680..169S} Shen, Y., Greene, J.~E.,
Strauss, M.~A., Richards, G.~T., \& Schneider, D.~P.\ 2008, \apj,
680, 169

\bibitem[Shields et al.(2003)]{2003ApJ...583..124S} Shields, G.~A.,
Gebhardt, K., Salviander, S., et al.\ 2003, \apj, 583, 124


\bibitem[Sikora
\& Madejski(2000)]{2000ApJ...534..109S} Sikora, M., \& Madejski, G.\
2000, \apj, 534, 109

\bibitem[Sironi
\& Spitkovsky(2011)]{2011ApJ...726...75S} Sironi, L., \& Spitkovsky, A.\ 2011, \apj, 726, 75

\bibitem[Soltan(1982)]{1982MNRAS.200..115S} Soltan, A.\ 1982, \mnras, 200,
115

\bibitem[Stawarz et al.(2007)]{2007ApJ...662..213S} Stawarz, {\L}., Cheung,
C.~C., Harris, D.~E., \& Ostrowski, M.\ 2007, \apj, 662, 213



\bibitem[Tavecchio et al.(2000)]{2000ApJ...544L..23T} Tavecchio, F.,
Maraschi, L., Sambruna, R.~M., \& Urry, C.~M.\ 2000, \apjl, 544, L23


\bibitem[Tchekhovskoy
\& McKinney(2012)]{2012MNRAS.tmpL.445T} Tchekhovskoy, A., \&
McKinney, J.~C.\ 2012, \mnras, 423, L55

\bibitem[Tchekhovskoy et al.(2011)]{2011MNRAS.418L..79T} Tchekhovskoy, A.,
Narayan, R., \& McKinney, J.~C.\ 2011, \mnras, 418, L79

\bibitem[Tremaine et al.(2002)]{2002ApJ...574..740T} Tremaine, S.,
Gebhardt, K., Bender, R., et al.\ 2002, \apj, 574, 740


\bibitem[Vitrishchak et al.(2008)]{2008MNRAS.391..124V} Vitrishchak, V.~M.,
Gabuzda, D.~C., Algaba, J.~C., et al.\ 2008, \mnras, 391, 124

\bibitem[Volonteri et al.(2007)]{2007ApJ...667..704V} Volonteri, M.,
Sikora, M., \& Lasota, J.-P.\ 2007, \apj, 667, 704


\bibitem[Wardle et al.(1998)]{1998Natur.395..457W} Wardle, J.~F.~C., Homan,
D.~C., Ojha, R., \& Roberts, D.~H.\ 1998, \nat, 395, 457

\bibitem[Woo
\& Urry(2002)]{2002ApJ...579..530W} Woo, J.-H., \& Urry, C.~M.\
2002, \apj, 579, 530

\bibitem[Woo et al.(2005)]{2005ApJ...631..762W} Woo, J.-H., Urry, C.~M.,
van der Marel, R.~P., Lira, P., \& Maza, J.\ 2005, \apj, 631, 762


\bibitem[Wu(2009)]{2009MNRAS.398.1905W} Wu, Q.\ 2009, \mnras, 398, 1905

\bibitem[Wu et
al.(2002)]{2002A&A...389..742W} Wu, X.-B., Liu, F.~K., \& Zhang,
T.~Z.\ 2002, \aap, 389, 742


\bibitem[Zhou
\& Cao(2009)]{2009RAA.....9..293Z} Zhou, M., \& Cao, X.-W.\ 2009,
Research in Astronomy and Astrophysics, 9, 293




\end{thebibliography}
\end{document}